\documentclass[epj]{svjour}
\usepackage{graphics}
\usepackage{multirow}
\RequirePackage{lineno} 
\newcommand{\sn}{${\rm \sqrt{s_{_{NN}}}}$}
\newcommand{\snn}{${\rm \sqrt{s_{NN}}}$ =}
\newcommand{\Au}{${\rm Au + Au}$}
\newcommand{\dAu}{${\rm d + Au}$}
\newcommand{\p}{${\rm p + p}$}
\newcommand{\Cu}{${\rm Cu + Cu}$}
\newcommand{\Pb}{${\rm Pb + Pb}$}
\newcommand{\pt}{${\rm p_{_{T}}}$}
\newcommand{\pta}{${\rm p^{(a)}_{_{T}}}$}
\newcommand{\ptt}{${\rm p^{(t)}_{_{T}}}$}
\def\dnch{${\rm dN_{ch}/d{\rm \eta}}$}
\def\dnchmid{${\rm dN_{ch}/d\eta|_{|\eta| <1}/\langle N_{part}/2 \rangle}$}
\def\dnchNp{${\rm dN_{ch}/ \langle N_{part}/2 \rangle}$}

\def\avgNp{${\rm \langle N_{part} \rangle}$}
\def\Np{${\rm N_{part}}$}

\def\FigureOne{
\begin{figure*}
\begin{center}
\resizebox{1.0\textwidth}{!}{\includegraphics{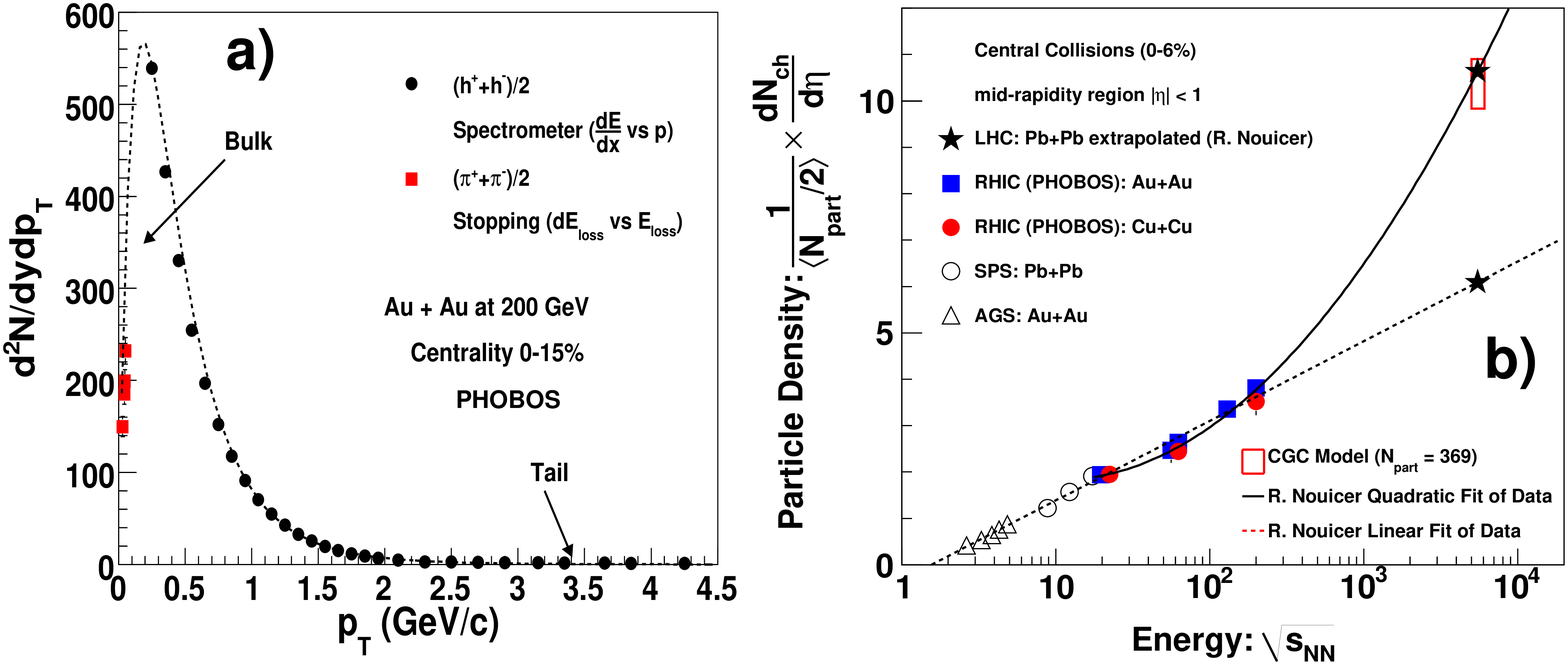}}
\caption{ Panel a) The measured distribution of charged-hadron as a
  function of transverse momentum \pt for the 15$\%$ most central
  \Au\ collisions at \snn\ 200 GeV. Panel b) The measured scaled
  pseudorapidity density ${\rm dN_{ch}/d\eta /\langle
    {1\over2}N_{part}\rangle}$ for ${\rm \mid \eta \mid < 1}$ in
  central Au+Au collisions at the AGS, in central \Au\ and \Cu\
  collisions at the RHIC and in central ${\rm Pb+Pb}$ collisions at
  the SPS
  \cite{RachidMoriond2002,AuAufrag,RachidPanic2006,CuCu2008}. The star
  symbols denote the extrapolation for ${\rm dN_{ch}/d\eta /\langle
    {1\over2}N_{part}\rangle}$ in central \Pb\ collisions at 5.5 TeV
  (LHC). The continuous curve corresponds to the logarithmic quadratic
  fit of the RHIC data: ${\rm f^{Q}_{AA} = 3.09 - 1.06 ln
    (\sqrt{s_{_{NN}}})+0.22 (ln (\sqrt{s_{_{NN}}}))^{2}}$. The dashed
  line represents the logarithmic linear fit of the AGS, SPS and RHIC
  data: ${\rm f^{L}_{AA} = 0.4749 + 0.77\ ln (\sqrt{s_{_{NN}}})}$. The
  error bars show the systematic errors.  The Color Glass Condensate
  model (CGC model) prediction for LHC is also illustrated as a red box
  \cite{Dima}.}
\label{fig1} 
\end{center}
\end{figure*}
}
\def\FigureTwo{
\begin{figure*}
\begin{center}
\resizebox{1.0\textwidth}{!}{\includegraphics{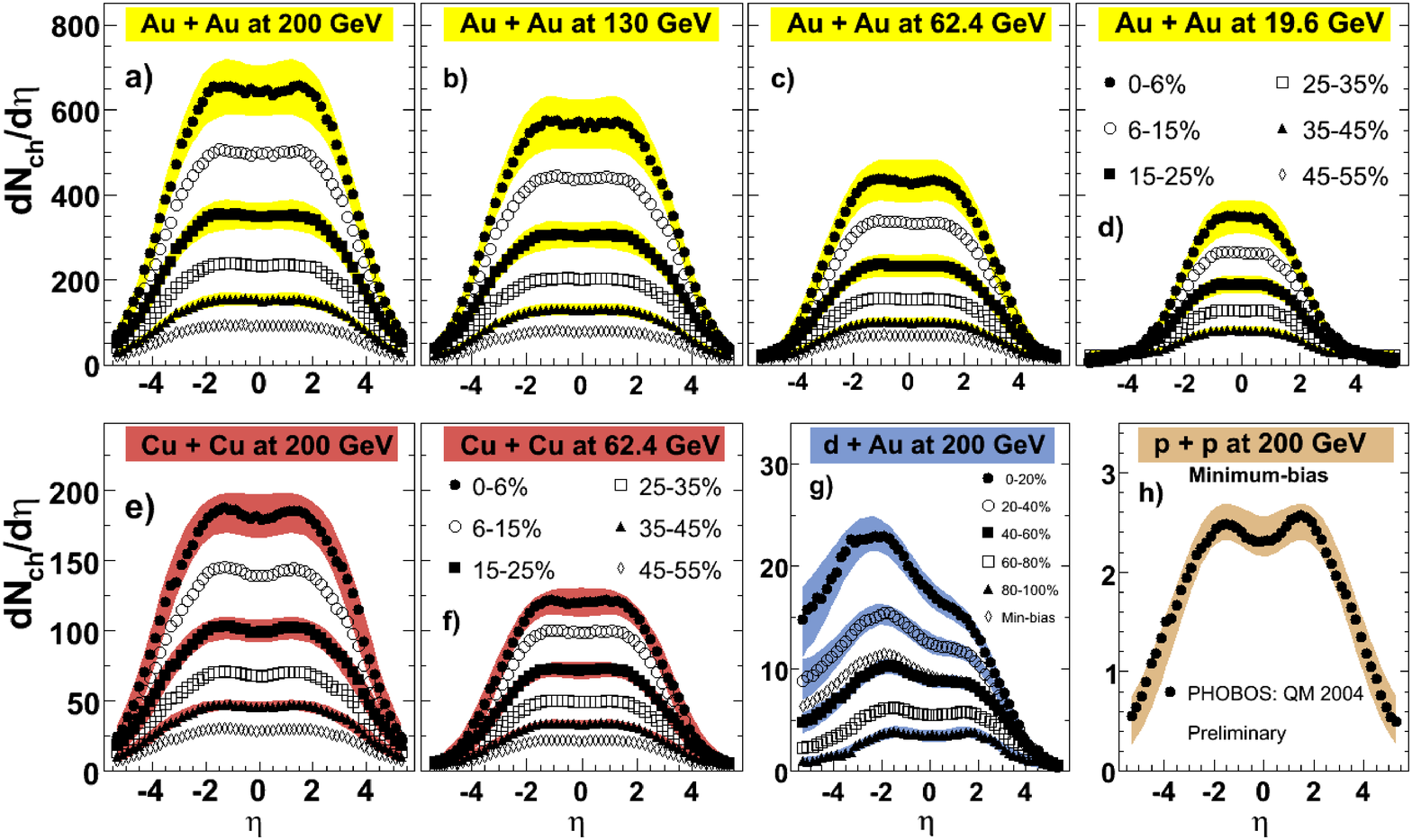}}
\caption{The measured distributions of pseudorapidity density in \Au,
  \Cu, \dAu, and \p\ collisions at RHIC energies
  \cite{RachidMoriond2002,AuAufrag,RachidPanic2006,CuCu2008}. The
  \dnch\ distributions for \Au, \Cu\ and \dAu\ collisions are plotted
  as a function of collision centrality. Typical systematic errors
  (90\% C.L.) are represented as bands for selected centrality
  bins. Statistical errors are negligible.}
\label{fig2} 
\end{center}
\end{figure*}
}
\def\FigureThree{
\begin{figure*}
\begin{center}
\resizebox{0.85\textwidth}{!}{\includegraphics{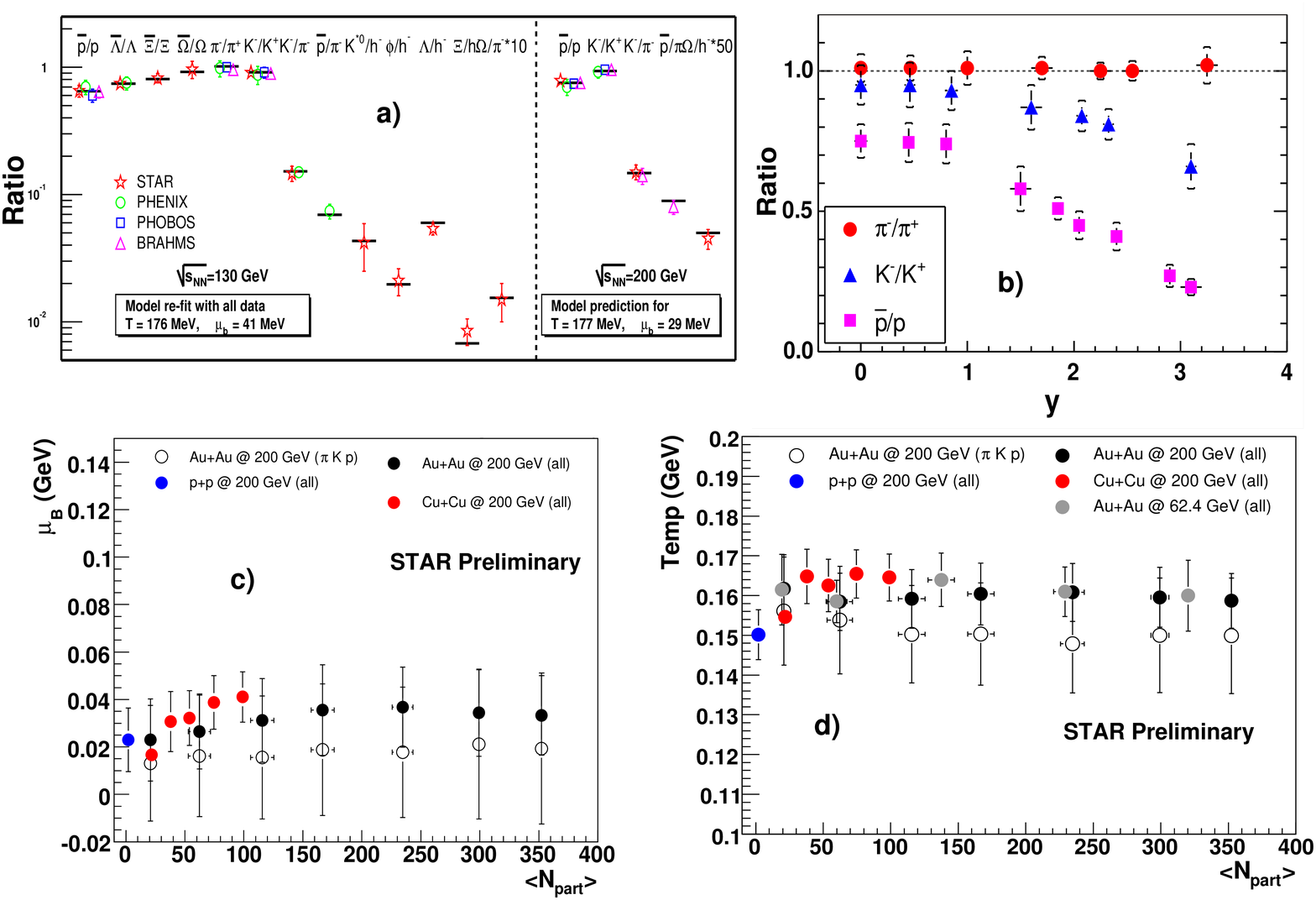}}
\caption{Panel a) compares the experimental data from different
  particle multiplicity ratios obtained in \Au\ collisions at RHIC
  energies at \snn\ 130 and 200 GeV with calculations from the thermal
  model \cite{Munzinger}. Panel b) shows the ratios of antiparticles to
  particles as a function of rapidity \cite{Brahms2003}. The errors bars depict the
  statistical errors and the caps indicate the combined statistical
  errors. Panels c) and d), respectively, illustrate the thermal model fit baryon
  chemical potential (${\rm \mu_{_{B}}}$) and the temperature chemical
  freeze-out (${\rm T_{ch}}$) as function of collisions centrality
  (\avgNp) \cite{Star2008}. ${\rm \mu_{_{B}}}$ and ${\rm T_{ch}}$ are obtained from
  \Au, \Cu\ and \p\ collisions at RHIC energies and for different
  particle species.}
\label{fig3} 
\end{center}
\end{figure*}
}
\def\FigureFour{
\begin{figure*}
\begin{center}
\resizebox{0.83\textwidth}{!}{\includegraphics{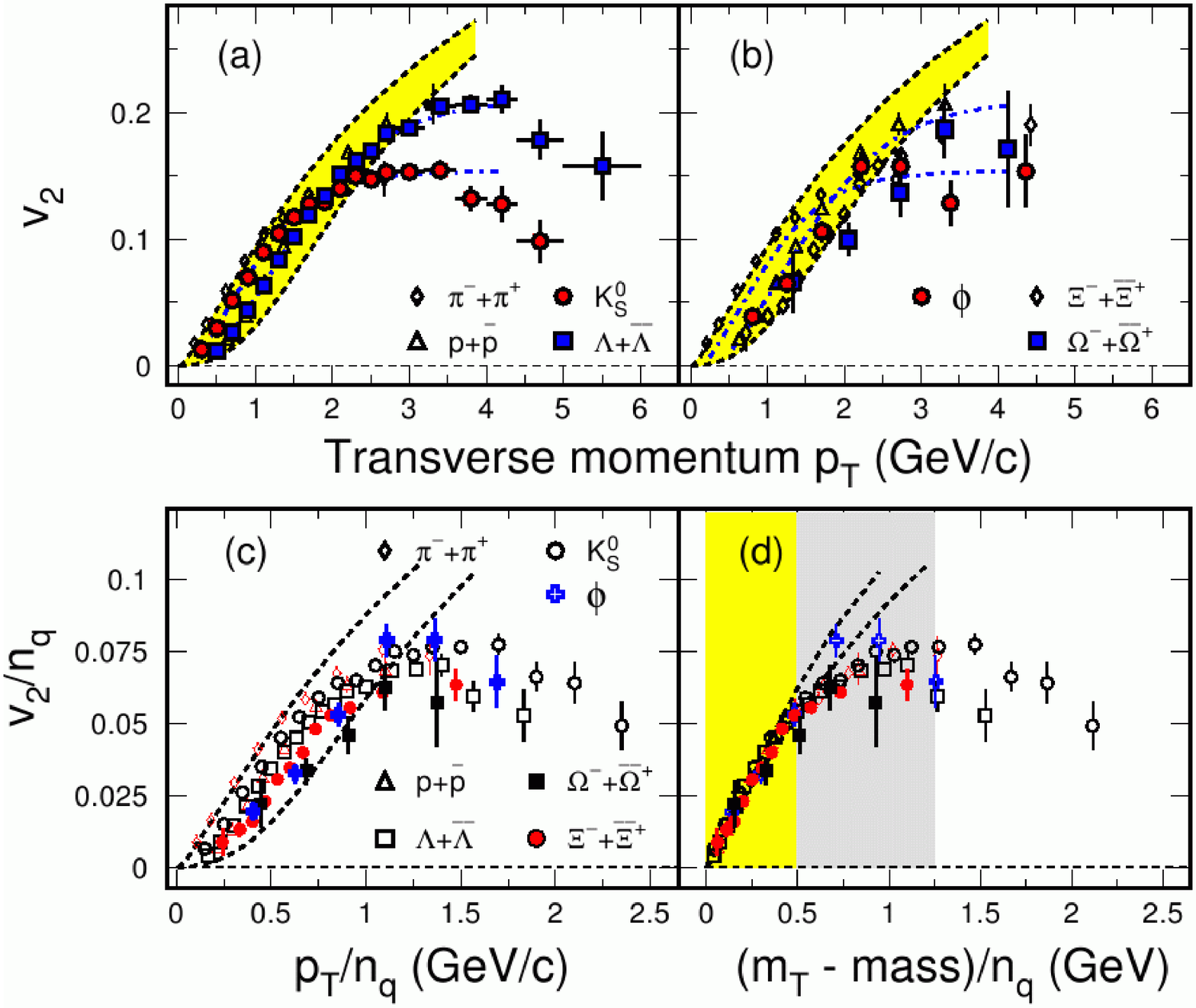}}
\caption{Identified hadron anisotropy: Panels a) and b) as a function
  of transverse momentum \pt, panel c) as a function of scaled
  \pt/${\rm n_{q}}$ and, panel d) as a function of scaled transverse
  kinetic energy (${\rm m_{T}}$ - mass)/${\rm n_{q}}$.  n$_{q}$ is the number of
  quarks in a given hadron (for mesons, ${\rm n_{q}}$ $=$ 2; and, for
  baryons: ${\rm n_{q}}$ $=$ 3). All data are from minimum-biased \Au\
  collisions at \snn\ 200 GeV \cite{NuXu}.}
\label{fig4} 
\end{center}
\end{figure*}
}
\def\FigureFive{
\begin{figure*}
\begin{center}
\resizebox{1.\textwidth}{!}{\includegraphics{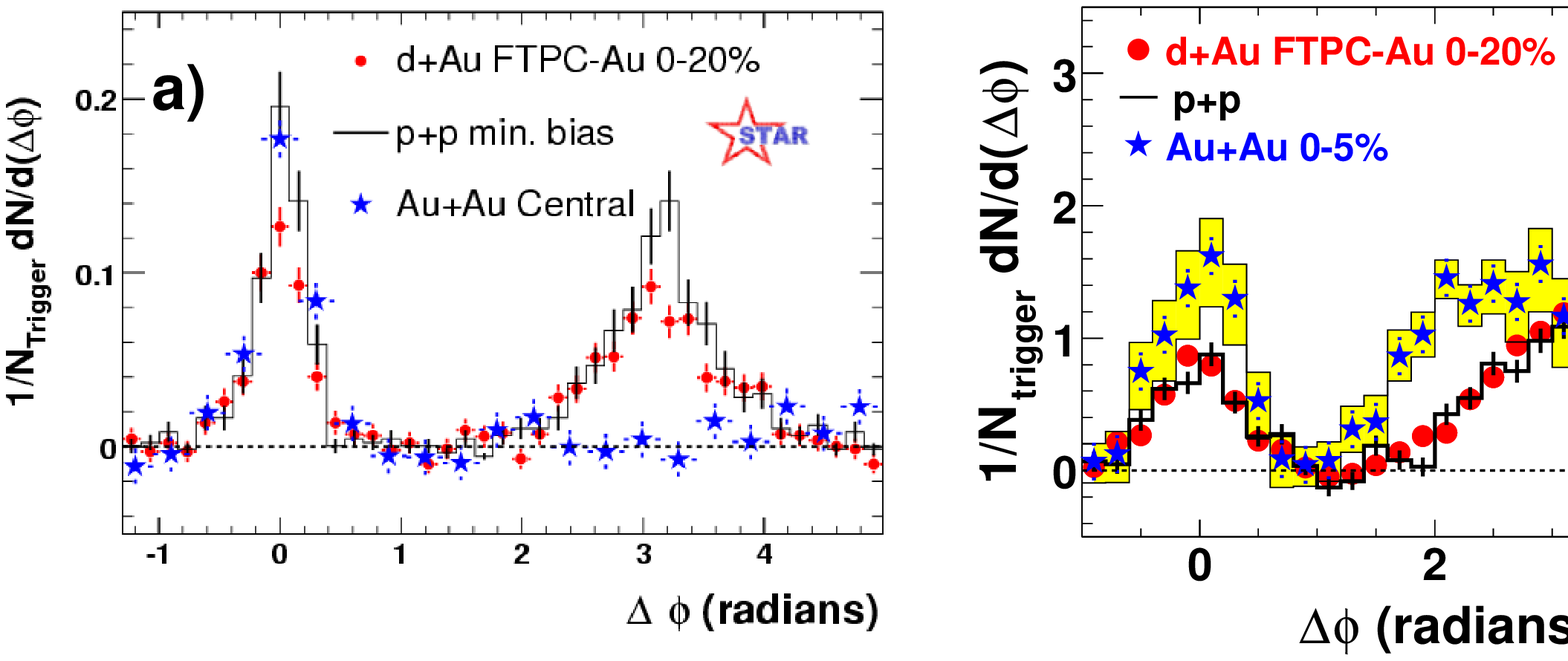}}
\caption{Measurements of two-particle angular correlations in \Au,
  \dAu\ and \p\ collisions at \snn\ 200 GeV in the presence of a
  trigger particle with \ptt under the condition 4 $<$ \ptt$<$ 6 GeV/c,
  and an associated particle with \pta: panel a) for 2 $<$\pta$<$\ptt
  GeV/c, and panel b) for 0.15 $<$\pta$<$ 2 GeV/c. The non-correlated
  background and the flow background were subtracted
  \cite{STARdi-jets}.}
\label{fig5} 
\end{center}
\end{figure*}
}
\def\FigureSix{
\begin{figure*}
\begin{center}
\resizebox{0.7\textwidth}{!}{\includegraphics{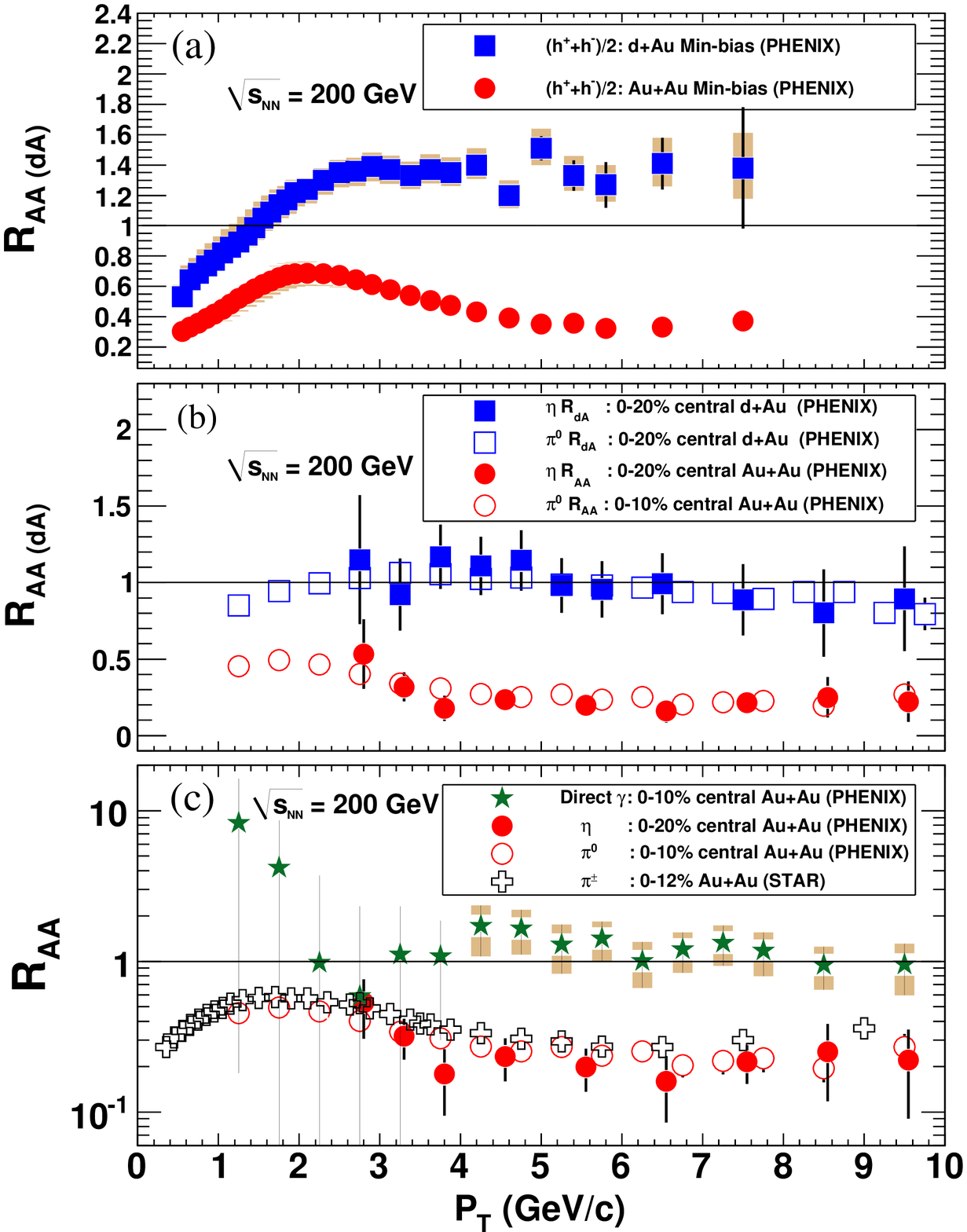}}
\caption{ Compilation of data of nuclear modification factor ${\rm R_{AA}}$
  from \Au\ collisions and ${\rm R_{dA}}$ of \dAu\ collisions at \snn\ 200
  GeV using Refs. \cite{CompilationRaa}. Panel a) compares ${\rm R_{AA}}$ 
  with ${\rm R_{dA}}$ for charged hadrons, (h$^{+} +$ h$^{-}$)/2, from
  minimum bias \Au\ and \dAu\ collisions. Panel b) compares ${\rm R_{AA}}$
  with ${\rm R_{dA}}$ for $\pi^{0}$ and $\eta$ from central \Au\
  and \dAu\ collisions. Panel c) compares the ${\rm R_{AA}}$ of Direct $\gamma$
  with ${\rm R_{AA}}$ of $\pi^{0}$, $\eta$ and $\pi^{\pi}$ in central
  \Au\ collisions. The error bars correspond to the statistical
  errors. For clarity, the systematic errors are shown as vertical
  bands.}
\label{fig6} 
\end{center}
\end{figure*}
}
\def\FigureSeven{
\begin{figure*}
\begin{center}
\resizebox{0.7\textwidth}{!}{\includegraphics{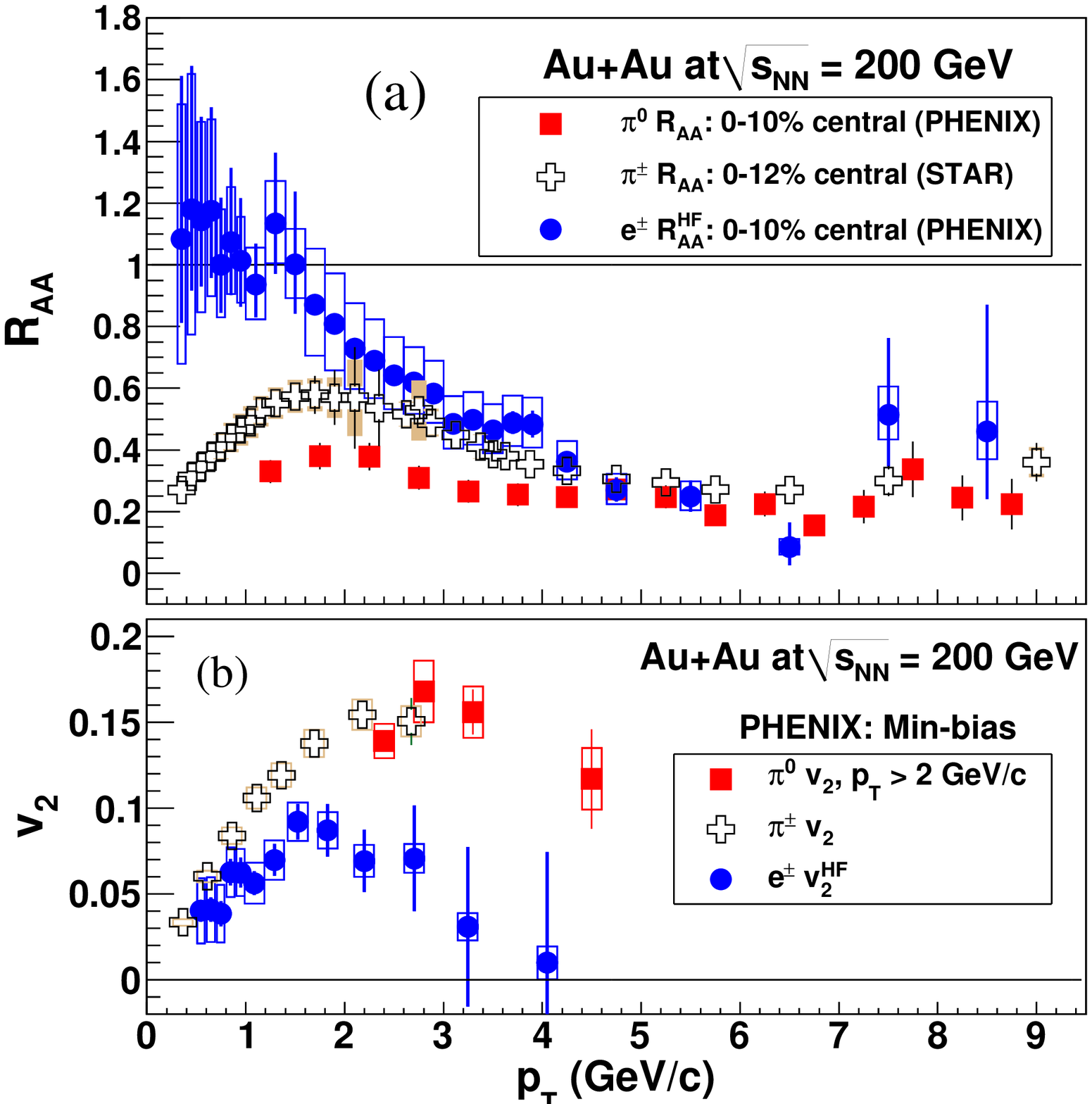}}
\caption{ Compilation of data on \Au\ collisions at \snn\ 200 GeV
  from Refs. \cite{CompilationRaa,HeavyPHENIX}. Panel a) represents the nuclear
  modification factor of heavy-flavor electrons ${\rm R_{AA}^{HF}}$ compared
  with the ${\rm R_{AA}}$ of $\pi^{0}$ and $\pi^{\pm}$ in central \Au\
  collisions. Panel b) considers the anisotropic flow of heavy-flavor
  electrons v$_{2}^{HF}$ with that of v$_{2}$ of $\pi^{0}$ and
  $\pi^{\pm}$ in minimum-bias \Au\ collisions \snn\ 200 GeV. The error
  bars correspond to the statistical errors. For clarity, the systematic
  errors are shown as vertical boxes and bands.}
\label{fig7} 
\end{center}
\end{figure*}
}
\begin{document}
\title{Formation of Dense Partonic Matter in High Energy Heavy-Ion Collisions: 
Highlights of RHIC Results} 
\author{Rachid Nouicer}
\offprints{\email{rachid.nouicer@bnl.gov}}
\institute{Brookhaven National Laboratory, Physics Department, Building 510D\\
        Upton, New York 11973-5000, U.S.A.}
\date{Received: date / Revised version: date}
\abstract{I review some important results from RHIC experiments. They
  were obtained in a unique environment for studying QCD bulk matter
  at temperatures and densities that surpass the limits where hadrons
  can exist as individual entities, raising the quark-gluon degrees of
  freedom to prominence. These findings support the major experimental
  observations from measuring the bulk properties of particle
  production, particle ratios and chemical freeze-out conditions,
  elliptic flow followed by hard probes measurements: di-jet fragment
  azimuthal correlations, high-\pt hadron suppression, and
  heavy-flavors probes. I present the measurements as a function of
  collision centrality, energy, system size and for different particle
  species. These results reveal that a dense strongly
  interacting medium was created in central \Au\ collisions at \snn\ 200 GeV:
   the RHIC discovery. Further, they suggest that this medium is
  partonic. However, the discoveries so far observed at RHIC are far
  from being understood fully. Accordingly, the focus of the experiments
  has shifted from the discovery phase to the detailed exploration
  phase of the properties of this medium.
} 
\authorrunning{R. Nouicer}
\titlerunning{Formation of Dense Partonic Matter in Heavy Ion Collisions ``Highlights of RHIC Results''}
\maketitle
\FigureOne%
\section{Physics motivation and RHIC achievements}
Quantum Chromodynamics (QCD) is considered the fundamental theory for
strong interactions. According to it, hadronic matter under extreme
dense and hot conditions must undergo a phase
transition~\cite{Shuryak80} to form a Quark Gluon Plasma (QGP) in
which quarks and gluons no longer are confined to the size of a
hadron. Recent results from lattice QCD at finite
temperature~\cite{Satz2003,Karsh2002} reveal a rapid increase in the
number of degrees of freedom associated with this deconfinement of
quarks and gluons from the hadronic chains. The transition point is at
a temperature ${\rm T \approx 170\ MeV}$ and energy density ${\rm
  \epsilon \approx 1\ GeV/fm^{3}}$. Under the same conditions, chiral
symmetry is restored~\cite{Satz2003}. Therefore, experiments search
for signatures of both QGP formation and the in-medium effects of
hadron properties. It was proposed that the required high densities
could be achieved via relativistic heavy ion
collisions~\cite{Greiner1975}.

Under RHIC (Relativistic Heavy Ion Collider) project, an
accelerator was constructed at Brookhaven National Laboratory (BNL)
from 1991 to 1999. RHIC was designed as a heavy-ion machine, able
to support the collision of a wide range of nuclei over a large range
of energies. Already, gold--gold, copper-copper and deuteron--gold
collisions have been attained at energies from \snn\ 19.6 to 200
GeV. Further, a polarized capability was added to RHIC allowing
transverse and longitudinal polarized protons to collide at energies
from \snn\ 200 to 410 GeV. By now, RHIC has made a major physics
discovery, namely the creation in high-energy central gold-gold
collisions of a new form of matter; dense and strongly interacting,
called the strongly coupled quark-gluon plasma, or sQGP.  This finding was
rated the top physics story of 2005 and the four experiments at RHIC,
BRAHMS, PHENIX, PHOBOS, and STAR, published, in white papers, evidence
of the existence of this new form of matter \cite{RHIC}. RHIC
accelerator also met and surpassed its specifications; namely, it
attained its energy goals, and exceeded the heavy-ion luminosity goals
by factor of 2, and polarized proton luminosity by a factor of 5.
Table~\ref{tab1} summarized the achievements of RHIC over the last
eight years.

In this article, I highlight the recent RHIC results underlying the
major experimental observations, i.e., ``discoveries''. I start with
discussions of bulk particle production and the initial conditions,
followed by particle ratios and chemical freeze-out conditions. I then
present our measurements of elliptic flow that are an indirect
signature of the existence of partonic matter, followed by hard-probe
measurements: di-jet fragment azimuthal correlations, high-\pt
hadron suppression, and heavy-flavors probes. These measurements of
hard probes afford direct signatures of highly interacting dense matter
created in central \Au\ collisions at \snn\ 200 GeV. The measurements,
for different particle species, are given as a function of collision
centrality, energy, and system sizes.
\section{Global properties of hadron production}
Measurements of charged-particle multiplicity and transverse energy
distributions, and of azimuthal anisotropies in heavy-ion collisions
afford information on the initial energy density and the entropy
production during the system's evolution, and are sensitive to a
variety of physics processes responsible for generating
multiparticles.  This knowledge is important for constraining model
predictions and indispensable for understanding and estimating the
accuracy of the more detailed measurements of, for example, jet or
quarkonia production. Figure~\ref{fig1}a shows the measured
distribution of charged-hadrons produced in the 15\% most central
\Au\ collisions at \snn\ 200 GeV~\cite{Rachid2003}. The distribution
illustrates a ``bulk'' and ``tail'' that are respectively related to
``soft'' and ``hard'' parton-parton scattering. We clearly observed
that the ``bulk'' constitutes the dominant part of charged-hadron
production. However, there is no clear separation between ``soft'' and
``hard'' processes. For this reason, an analysis undertaken as a
function of centrality, transverse momentum, energy, and system sizes
is required for better understanding particle production.
\FigureTwo%
\subsection{Charged-particle density distributions and initial conditions}
One important observable in heavy-ion interactions is the number of
charged-particles produced in a collision, called the charged-particle
pseudorapidity density \dnch. This number (\dnch) is proportional to
the entropy density at freeze-out and, since entropy cannot be
destroyed (even in non-equilibrium systems), pseudorapidity density
provides information on the initial state of parton density, and any
further entropy produced during subsequent evolution
\cite{RachidMoriond2002}.

Fig.~\ref{fig1}b shows the charged particle densities near the
mid-rapidity region, \dnchmid, where \avgNp\ is the average number
of nucleon participant pairs, for \Au\ (RHIC: 19.6, 62.4, 130 and 200
GeV), \Cu\ (RHIC: 22.4, 62.4 and 200 GeV), \Pb\ (SPS: 8.83, 12.28 and
17.26 GeV) and \Au\ (AGS: 2.63, 3.28, 3.84, 4.29, 4.86 GeV); they
present the 6\% most central collisions as a function of
center--of--mass collision energy (\sn)
\cite{RachidMoriond2002,AuAufrag,RachidPanic2006,CuCu2008}.  The
results in this figure (see Fig.~\ref{fig1}b) suggest that the
particle density rises approximately logarithmically with
energy. Comparing the finding for \Cu\ and \Au\ indicates that, for
the most central events in symmetric nucleus-nucleus collisions, the
particle density per nucleon participant-pair does not depend on the
size of the two colliding nuclei but only on the collision's
energy~\cite{RachidPanic2006,CuCu2008}.  

Based on the data presented in the Fig.~\ref{fig1}b, we can establish
two types of fits to predict (extrapolate) the results of \dnch\ at
the mid-rapidity region in \Pb\ collisions at the LHC energy 5.5 TeV:
\begin{enumerate}
\item The logarithmic quadratic fit of the RHIC data is shown as the solid
curve in Fig.~\ref{fig1}b and its fit function is:
\begin{equation}
{\rm f^{Q}_{AA} = 3.11 - 1.07 ln (\sqrt{s_{_{NN}}})+0.23 (ln
(\sqrt{s_{_{NN}}}))^{2}}
\end{equation}
this formulation allows us to extrapolate the scaled density per nucleon
participant pair for central \Pb\ collisions at LHC energy (5.5 TeV):
$$ {{\rm \frac{1}{\langle N_{part}/2\rangle} \times \frac{dN_{ch}}
    {d\eta} (Pb+Pb\ at\ 5.5\ TeV) = 10.95}}$$ 
Using a Glauber model calculation for the 6\% most central \Pb\ collisions at 5.5 TeV (the
total inelastic cross section used in the Glauber model calculation is
$\sigma_{_{NN}}$ = 64 mb), we obtained the value of \avgNp =
365.5, from which we can deduce the value of the unscaled pseudorapidity
density at the mid-rapidity region:
$${{\rm \frac{dN_{ch}} {d\eta} (central: 0-6\%, Pb + Pb\ at\ 5.5\ TeV) = 2002}}$$
\item The logarithmic linear fit of the data from the AGS, SPS and
  RHIC is shown by the dashed line in Fig.~\ref{fig1}b, and has the fit
  function:
\begin{equation}
{\rm f^{L}_{AA} = -0.33 + 0.75\ ln (\sqrt{s_{_{NN}}})}
\end{equation}
With this fit equation, we extrapolated the scaled particle density per
nucleon participant pair at the mid-rapidity region to LHC energy, \snn\
5.5 TeV:
$$ {{\rm \frac{1}{\langle N_{part}/2\rangle} \times \frac{dN_{ch}}
    {d\eta} (Pb +Pb\ at\ 5.5\ TeV) = 6.1}}$$ 
For \avgNp\ = 365.5 obtained via the Glauber model calculation (using $\sigma_{_{NN}}$ = 64
mb) for 6\% central \Pb\ collisions at 5.5 TeV, this yielded:
$${{\rm \frac{dN_{ch}} {d\eta} (central: 0-6\%, Pb + Pb\ at\ 5.5\ TeV) = 1120}}$$
\end{enumerate}
The prediction of the Color Glass Condensate model (CGC model) \cite{Dima} for
\dnchmid\ in \Pb\ collisions at LHC energy, \snn\ 5.5 TeV, 
agrees very well with the results of extrapolation obtained with the logarithmic
quadratic fit, as shown in Fig.~\ref{fig1}b.\FigureThree%

Fig.~\ref{fig2} plots the measured \dnch\ of primary charged
particles over a broad range of pseudorapidity, ${\rm |\eta| <5.4}$,
for \Au\ \cite{RachidMoriond2002,AuAufrag} and \Cu\
\cite{RachidPanic2006,CuCu2008} collisions under a variety of collision
centralities and RHIC energies. The \dAu\
\cite{RachidQM2004,dAuPRL,dAuPRC} and \p\ \cite{RachidQM2004} data at
RHIC energies also are shown. The \Cu, \Au\, \dAu\ and \p\ data at all
energies were obtained with the same detector setup in the PHOBOS
experiment \cite{PHONIM}. This is optimal because common systematic
errors cancel each other out in the ratio.  With this configuration,
we were able to examine comprehensively particle production in \Cu\
and \Au\ collisions for the same number of nucleon participant pairs,
for the same fraction of total inelastic cross sections, and for the
same geometry in both systems \cite{CuCu2008}. Both the height and
width of the \dnch\ distributions increase as a function of energy in
both systems.  Comparing the results revealed an interesting feature,
namely that the best agreement of \dnchNp\ distributions over the full
coverage, ${\rm |\eta| <5.4}$, for central and for peripheral \Cu\ and
\Au\ collisions is obtained for centrality bins selected to yield a
similar geometry, i.e., a comparable value of \Np/2A (where A is the
atomic number) in both systems \cite{CuCu2008}.

One question to be asked is whether mechanisms in the limiting
fragmentation region (i.e.~high ${\rm |\eta|}$) are distinct from
those at mid-rapidity (${\rm |\eta| < 1}$). From the results shown in
Fig.~\ref{fig2} for \Au, \Cu, \dAu, and \p\ collisions, there is no
obvious evidence for two separate regions at any of the RHIC energies
\cite{RachidPanic2006}. However, to date no such anomalies have been
noted, at the AGS, SPS, or RHIC. So far, all results on particle
pseudorapidity densities point to a rather smooth evolution in
centrality and \sn. This of course, does not necessarily imply the
absence of a phase transition, but rather, might indicate the
insensitivity of these observables to the early phase of the collision
and/or might suggest a second order phase transition (or a cross-over)
\cite{Thomas}. We note that charged-particle production mechanisms for
\dnch\ shown in Fig.~\ref{fig2} are mostly obtained from the ``soft''
processes (see Fig.~\ref{fig1}a).
\subsection{Particle ratios and chemical freeze-out conditions}
The chemical freeze-out point is the stage in the evolution of the
hadronic system when inelastic collisions cease and the relative
particle ratios become fixed; this point, is defined by the
temperature chemical freeze-out, ${\rm T_{_{ch}}}$, and also by the
baryon chemical potential, ${\rm \mu_{_{B}}}$. These parameters, ${\rm
  T_{_{ch}}}$, and ${\rm \mu_{_{B}}}$ determine the particle
composition of the hadronic final state.  After chemical freeze-out,
the particle composition inside the fireball is fixed, but elastic
collisions keep the system intact until the final, thermal
freeze-out. At this stage the momentum distributions of particles are
final and no longer change.  Therefore, the transverse momentum
spectra determine the parameters of the thermal freeze-out.
Thereafter, statistical interpretation of particle production becomes
an appropriate approach for evaluating heavy-ion collisions at high
energies; because large multiplicities of hadrons are created. We can
assume that the nuclear matter created in these collisions form an
ideal gas that can be characterized by a grand-canonical
ensemble. Using thermodynamic concepts to describe multiparticles
production has a long history~\cite{Hagedorn1965}. The concept of a
temperature applies only to systems in at least local thermal
equilibrium. The assumption of a locally thermalized source in
chemical equilibrium can be tested by applying statistical thermal
models to describe the ratios of various emitted particles.
\FigureFour%

Figure~\ref{fig3}a compares the RHIC's experimental particle ratios
and statistical thermal model calculations for \Au\ collisions at 130
and 200 GeV \cite{Munzinger}; note that the measurements were taken in
the mid-rapidity region $|\eta| < 1$. They demonstrate quantitatively
the high degree of equilibration achieved for hadron production in
central Au$+$Au collisions at RHIC energies. Values were obtained for
(${\rm T_{ch}}$, ${\rm \mu_{_{B}}}$) of (174 $\pm$7, 46$\pm$5) and
(177$\pm$7, 29$\pm$6) at 130 and 200 GeV, respectively, The model
matches well with these results, and there is no indication of a
significant change in ${\rm T_{_{ch}}}$ at the two energies, 130 and
200 GeV. There is a drop in $\mu_{_{B}}$ from $\sim$ 46
MeV, at \snn\ 130 GeV to 29 MeV at \snn\ 200 GeV. We note that ratios
involving multi-strange baryons are well reproduced, as is the
$\phi$/h$^{-}$ ratio. Even relatively wide resonances such as the
K$^{*}$'s fit well with the picture of a chemical freeze-out.

The BRAHMS collaboration reported interesting results on particle
ratios as a function of rapidity \cite{Brahms2003}.
Figure~\ref{fig3}b shows the $\pi^{-}/\pi^{+}$, $K^{-}/K^{+}$ and
$\bar{p}/p$ ratios as a function of this parameter. We observe that
the $\pi^{-}/\pi^{+}$ ratio is consistent with unity over the
considered rapidity range, while the $K^{-}/K^{+}$ ratios drops almost
by 30\% at y $=$ 3 from its mid-rapidity value, and the $\bar{p}/p$
ratio by almost 70\%. Fig.~\ref{fig3}d and c show thermal model fit
temperature and baryon potential at chemical freeze-out as a function
of collision centrality (\avgNp) in \Au, \Cu\ and \p\ collisions at
\snn\ 200 GeV \cite{Star2008}. Temperature seems constant as a
function of \avgNp\ and apparently independent of system size (\Au\
and \Cu\ systems). The baryon chemical potential exhibits small
variations as a function of \Np, consistent with of being independent
of system size in \Au\ and \Cu\ collisions.\FigureFive%
\FigureSix%
\section{Evidence of partonic matter at RHIC}
\subsection{Anisotropic flow of hadrons}
The anisotropic flow of hadrons has been studied extensively in
nucleus-nucleus collisions at the SPS and RHIC as a function of
pseudorapidity, centrality, transverse momentum and energy
\cite{RHIC,FlowSPS,RachidQM2006}. In non-central collisions of heavy
ions at high energy, the configuration space anisotropy is converted
into a momentum space anisotropy. The dynamics of the collision
determine the degree of this transformation. For a symmetric system,
like \Au, the second Fourier expansion is a good parameterization of
the anisotropy. At RHIC, a strong anisotropic flow (v$_{2}$) was
observed for all hadrons measured suggesting that a strongly
interacting system was created in the collisions.

Hydrodynamic models that assume the formation of a QGP were used to
model the behavior of the medium so created \cite{hydro}. Comparison
of the data obtained at mid-rapidity region presented by the STAR,
PHOBOS, and PHENIX collaborations \cite{RHIC,RachidQM2006} to the
hydrodynamic model \cite{hydro} affords strong evidence that the
originated medium behaves as a near ideal fluid.

Fig.~\ref{fig4} shows the anisotropic flow distributions, v$_{2}$, for
identified hadrons ${\rm \pi}$, ${\rm K^{0}_{S}}$, p, ${\rm \Lambda}$,
${\rm \Xi}$, ${\rm \Omega}$ and ${\rm \phi}$ \cite{NuXu}. Calculations
from a hydrodynamic model are depicted in dashed lines, while the
dot-dashed lines represent the results of fit for quark number scaling
${\rm n_{q}}$ (for mesons, ${\rm n_{q}}$ $=$ 2; and, for baryons:
${\rm n_{q}}$ $=$ 3). In the lower \pt region, \pt $\le$ 2 GeV/c, the
value of v$_{2}$ is inversely related to the mass of the hadron, that
is, characteristic of hydrodynamic collective motion in operation, see
Fig~\ref{fig4}a and b. At the intermediate \pt region, the dependence
is different. Instead of a mass dependence, there seems to be a hadron
type dependence, Fig.~\ref{fig4}(b).  An interesting result concerns
the $\phi$ meson, whose mass is close to that of p and ${\rm
  \Lambda}$. The v$_{2}$ distribution illustrated as circles in
Fig.~\ref{fig4}b, is the same as the v$_{2}$ for $\pi$ and ${\rm
  \Omega}$. It is known that the $\phi$ meson does not participate as
strongly as others do in hadronic interactions, nor can $\phi$ mesons
be formed via the coalescence-like $K^{+} + K^{-}$ process in high
energy collisions \cite{StarKprocess}. Hence, the strong v$_{2}$ we
recorded must have developed before hadronization. This observation,
together with the $\Omega$ v$_{2}$ results offer clear evidence for
partonic collectivity \cite{RHICStar}. To demonstrate the scaling
properties of v$_{2}$, the following transformation have been
performed. The measured v$_{2}$ was scaled by the number of valence
quarks in a given hadron. For mesons and baryons, respectively, they
are ${\rm n_{q}}$ $=$ 2 and 3. The \pt also was scaled with the same
${\rm n_{q}}$. The results are shown in Fig.~\ref{fig4}c. To include
the effect of the collective motion, the \pt was transfered to the
transverse kinetic energy ${\rm KE_{T}}$ $\equiv$ ${\rm m_{T}-}$mass
$=$ ${\rm \sqrt{p^{2}_{T}-m^{2}}-m}$ scaled by ${\rm n_{q}}$, where m
mass of the particle. The outcome is depicted in Fig.~\ref{fig4}d. All
of the hadron v$_{2}$ scaled nicely up to (${\rm m_{T}}$ $-$
mass)/${\rm n_{q}}$ $\sim$ 1.2. Hydrodynamic calculations \cite{Huo},
shown as dashed-lines in the plot, also are scaled in the low \pt
region. At high-\pt, the values of v$_{2}$ appear to fall.  These
observations about scaling reveal that before hadronization, quarks
already have acquired the collective motion v$_{2}$, and that when
they coalesce, the v$_{2}$ is passed to newly formed hadrons. Again
this observation verifies partonic collectivity.
\section{Signatures of dense matter at RHIC}
\subsection{Di-jet fragment azimuthal correlations}
The study of high transverse momentum hadron production in
heavy ion collisions at RHIC provides an experimental probe of the QCD
matter in the most dense stage of the collisions, wherein
quark-gluon deconfinement is likely to occur \cite{Jacobs2005}. In particular,
two-hadron azimuthal correlations support the assessment of
back-to-back, hard-scattered partons that propagate in the medium
before fragmenting into jets of hadrons, thereby serving as a
tomography probe of the medium. 

Fig.~\ref{fig5} shows measurements of two-particle correlations
\cite{STARdi-jets} given a trigger particle \ptt under the condition
4$<$ \ptt$<$6 GeV/c, along with an associated particle
with \pta: 1) for 2 $<$ \pta$<$ \ptt GeV/c as
shown in Fig.~\ref{fig5}a, and 2) for 0.15 $<$ \pta$<$ 2
GeV/c as shown in Fig.~\ref{fig5}b. The two-particle correlation are
presented as a function of the difference between the azimuthal angles
of the two particles (${\rm \Delta \phi}$) produced in \Au, \dAu\ and
\p\ collisions at \snn\ 200 GeV. These measurements were made at the
mid-rapidity region ${\rm 0<|\Delta\eta|<1.4}$

Fig.~\ref{fig5}a demonstrates that the trigger-side correlation peak in
central \Au\ collisions apparently is the same as that measured in \p\
and \dAu\ collisions but the away-side jet correlation in \Au\ has
vanished. This observation is consistent with a large energy loss in the
medium causing it to be opaque to the propagation of high momentum
partons. However, it is noticeable in Fig.~\ref{fig5}b, which has a
wide \pta, 0.15 $<$ \pta $<$ 2 GeV/c, that the away side jets have
not really disappear, but simply lost energy so that the away-side
correlation peak has became much wider than that of the \p\ and \dAu\
collisions. These collective measurements, from \p, \dAu\ and \Au\ of
two-particle correlations, point to the creation of a dense medium in
central \Au\ collisions at \snn\ 200 GeV.
\subsection{High-\pt hadron suppression: jet-suppression}
In heavy ion collisions (from AGS to RHIC energies) hadron production
at the mid-rapidity region (${\rm |y| <}$ 1.5) rises with increasing
collision energy. At RHIC, the central zone is almost baryon
free~\cite{BrahmsBaryon}. Particle production is large and dominated
by pair production, and the energy density seems to exceed
significantly that required for QGP formation \cite{RHIC}.

Recently, RHIC experiments revealed suppression of the high transverse
momentum component, \pt of hadron spectra at the mid-rapidity region
in central Au+Au collisions compared to scaled momentum spectra from
p+p collisions at the same energy, \snn\ 200 GeV \cite{RHIC}. This
effect, originally proposed by Bjorken, Gyulassy and others
\cite{Bjorken} rests on the expectation of a large energy loss of high
momentum partons scattered in the initial stages of collisions in a
medium with a high density of free color charges. According to QCD
theory, colored objects may lose energy by the bremsstrahlung
radiation of gluons \cite{Gaard}. Such a mechanism would strongly
degrade the energy of leading partons, reflected in the reduced
transverse momentum of leading particles in the jets emerging after
fragmentation into hadrons. The STAR experiment established that the
topology of high-\pt hadron emission is consistent with jet emission,
so that jet-suppression is a valid concept.

Fig.~\ref{fig6} shows compilation of data using
Refs. \cite{CompilationRaa} for the nuclear modification factors
measured for different collisions systems at RHIC energies. The
nuclear modification factor is defined as:
$${\rm  R_{AA} (p_{T})= \frac{yield\ per\ A+A\ collisions}{N_{bin} \times(yield\ per\ p+p\ collisions) }}$$
$${\rm = \frac{d^{2}N^{^{A+A}}/dp_{T}d\eta}{N_{bin}\
  d^{2}N^{^{p+p}}/dp_{T}d\eta}}$$  
It involves scaling measured distributions of nucleon-nucleon transverse 
momentum by the number of expected incoherent binary collisions, 
${N_{bin}}$ \cite{RAAdefinition}. In the absence of any modifications due to the 
`embedding' of elementary collisions in a nuclear collision, 
we expect ${\rm R_{AA}}$ = 1 at high-\pt. At low \pt, where particle 
production follows a scaling with the number of participants, the above 
definition of ${\rm R_{AA}}$ leads to ${\rm R_{AA}}$ $<$ 1 for \pt $<$ 2 GeV/c.
 
In \dAu\ collisions at RHIC energy (\snn\ 200 GeV) as shown in
Fig.~\ref{fig6}a, the ${\rm R_{dA}}$ for charged hadrons (h$^{+}$ $+$
h$^{-}$)/2 is enhanced. This enhancement was expected as a consequence
of Cronin enhancements, now understood as an initial state effect
\cite{Accardi} which also is seen in pA collisions.  
The Cronin effect is associated with the initial multiple scattering
of high momentum partons. In contrast, RHIC experiments discovered a
factor of 4--5 suppression, in central \Au\ (at \snn\ 200 GeV). In the
same context, the ${\rm R_{dAu}}$ of $\pi^{0}$ and $\eta$ in central
\dAu\ collisions exhibits no suppression of high-\pt, in contrast
with the ${\rm R_{AA}}$ of $\pi^{0}$ and $\eta$ in central \Au\
collisions that is suppressed, as shown in Fig.~\ref{fig6}b. At \pt
$\sim$ 4 GeV/c, we find a ratio ${\rm R_{dAu}}$/${\rm R_{AA}}$
$\approx$ 5. Indeed, the ${\rm R_{dA}}$ distribution shows the Cronin
type enhancement observed at lower energies as in \Pb\ collisions at
17.3 GeV/c \cite{Gyulassy2005}.

Fig.~\ref{fig6}c summarizes the present status of ${\rm R_{AA}}$ for
direct photons, ${\rm \pi^{0}}$ and ${\rm \eta}$ in central \Au\
collisions at \snn\ 200 GeV. The ${\rm R_{AA}}$ for direct photons are
not suppressed because they do not interact strongly with the
medium. The ${\rm R_{AA}}$ for both ${\rm \pi^{0}}$ and ${\rm \eta}$'s
mesons exhibit the same suppression relative to the point-like scaled
\p\ data by a factor of $\sim$ 5 which appears to be constant for \pt
$>$ 4 GeV/c while the ${\rm \eta}$ mass is much larger than that of
${\rm \pi^{0}}$. This observation combined with Fig.~\ref{fig4} is
clearly an indication of partonic nature of suppression.

This data of ${\rm R_{AA}}$ were described by theoretical calculations of
parton energy loss in the matter created in Au+Au collisions
\cite{Vitev1}. From these theoretical frameworks, we have learned that
the gluon density ${\rm dN_{g}}$/dy must be approximately 1000
\cite{Vitev2}, and the energy density of the matter created in the
most central collisions must be approximately 15 GeV/fm$^{3}$ to
account for the large suppression observed in the data \cite{Vitev3}.

\FigureSeven%
\section{Heavy-flavors as probes for dense medium}
As discussed above, RHIC has produced in high-energy central
\Au\ collisions at \snn\ 200 GeV a new form of matter that is dense,
strongly interacting and can be partonic, called sQGP. This new
state of strongly interacting matter can closely resemble a ``perfect''
fluid, with very low viscosity and high opacity explained by the
equations of hydrodynamics \cite{hydro}.  The research focus now has shifted
from this initial discovery to a detailed exploration of partonic matter.
Particles carrying a heavy flavor, i.e., charm or beauty quarks, are
powerful tool for studying the properties of the hot, dense medium
created in high-energy nuclear collisions: they are generated early in
the reaction, and subsequently diffuse in the putative sQGP.

The study of heavy flavor in relativistic nuclear collisions follows
two different approaches: 1) The direct reconstruction of the heavy
flavor meson, and 2) the identification of electrons and muons from
semi-leptonic decays of such mesons. The PHENIX and STAR experiments
at RHIC explore one or both of these methods. The STAR collaboration
is directly reconstructing heavy flavor mesons the decay channel ${
  D^{0} \rightarrow K^{-} + \pi^{+}}$ in \dAu\ and \Au\
collisions. Employing semi-leptonic decays of heavy flavor mesons,
such as ${ D^{0} \rightarrow\ e^{+} + K^{+} + \nu}$ over a broad
\pt range, more efficiently measures charm and bottom quark
production and overcomes the limitation imposed by directly
reconstructing such mesons. 

At RHIC, two methods are utilized to measure heavy flavor production
via semi-leptonic decays: 1) Identification of electrons from the
decays of ${D}$ and ${B-}$mesons, and 2) identification of muons from
${D-}$meson decays. The PHENIX and STAR experiments pursue the
analysis of non-photonic electrons (from heavy flavor)
\cite{HeavyPHENIX,HeavySTAR}. Electron identification in PHENIX
largely is based on using the Ring Imaging Chernkov detector (RICH) in
conjunction with a highly granular EMC (electromagnetic
calorimeter). Their momentum is derived from the curvature of the
track (due to a magnetic field up to 1.15 Tesla) reconstructed from
drift and pad chambers. STAR identifies of electrons is done using
information on ${dE/dx}$ and momentum gleaned from the Time Projection
Chamber (TPC) and the Time of Flight (ToF) data for the low to
moderate \pt (\pt $<$ 4-5\ GeV/c) electrons. The barrel
electromagnetic calorimeter's (EMCAL's) data is used for moderate to
high-\pt (\pt $>$ 1.5 GeV/c) electrons \cite{HeavySTAR}. A major
difficulty in both collaborations' electron analyzes is the fact that
there are many sources of electrons other than the semi-leptonic
decays of heavy flavor mesons. The main sources of background are
photon conversion in the detector material (less significant in PHENIX
that has a lesser amount of material than STAR), and $\pi^{0}$ and
$\eta$ Dalitz decays. Other sources of background such as $\omega$,
$\phi$ and $\rho$ decays also must be taken into account. The
background sources usually are called photonic electrons.

Fig.~\ref{fig7} shows compilation of data using
Refs. \cite{CompilationRaa,HeavyPHENIX,HeavySTAR} of the nuclear modification
factors and elliptic flow measured for \Au\ collisions systems at \snn
200 GeV. Fig.~\ref{fig7}a compares the nuclear modification factor
R$_{AA}$ of heavy flavor electrons to $\pi^{0}$ data, and to
$\pi^{\pm}$ data obtained from central \Au\ collisions at 
\snn\ 200 GeV. We observe clear suppression of heavy flavor electrons in
central events in high-\pt. For \pt $>$ 4 GeV/c, the R$_{AA}$
of heavy flavors is surprisingly similar to that for $\pi^{0}$ and
$\pi^{\pm}$.  Fig.~\ref{fig7}b illustrates the distribution of the
anisotropy of heavy flavors electrons (v${^{HF}_{2}}$) as a function
the particles' transverse momentum, \pt, in minimum bias \Au\
collisions at \snn\ 200 GeV. Collective behavior is apparent in
heavy-flavor electrons (v${^{HF}_{2}}$ $>$ 0); but however, it is
lower than v$_{2}$ of $\pi^{0}$ for \pt $>$ 2 GeV/c.

This data presented in Fig~\ref{fig7} indicates that heavy flavors
strongly coupled with the medium. The observed suppression at high-\pt
$>$ 4 GeV/c can be explained in terms energy loss from the
heavy-flavors into the medium. Ref. \cite{HeavyPHENIX} offers
different theoretical predictions for heavy-flavor electrons in
central \Au\ collisions, considering different energy loss
mechanisms. However, I note that in Fig.~\ref{fig7}b, the data extend
only to around \pt $\approx$ 4 GeV/c, and there are large
statistical errors in the v${^{HF}_{2}}$. This implies that we can
reach a comprehensive conclusion about R${^{HF}_{AA}}$ and
v${^{HF}_{2}}$ only when we have more data on \Au\ collisions at \snn\
200 GeV is needed and an upgrade of PHENIX and STAR detectors is
required. Furthermore, we need measurements of the particles
identified as decay products from charm or beauty flavors using the
displacement of their trajectories from the collision vertex
(measuring the distance of closest approach, DCA). Fortunately, both
detectors PHENIX and STAR already were upgrade by building silicon
vertex trackers for heavy flavors \cite{RachidNIM,STARNIM}.
\section{Conclusions}
I discussed measurements from RHIC collisions on bulk particle
production, the initial conditions of the collisions, particle ratios
and chemical freeze-out conditions, flow anisotropy of particles,
di-jet fragment azimuthal correlations, high-p$_{t}$ hadron
suppression and heavy-flavors probes as a function of collision
centrality, energy, system sizes and for different particle species.
The measurements suggest that RHIC discovered a new state of matter in
central \Au\ collisions at \snn\ 200 GeV. This new form of matter is
a hot, dense, strongly interacting and is called ``sQGP''. Whilst this
material, and its discovery, is far being fully understood, the focus
of the RHIC experiments has shifted from the discovery phase to a
detailed exploration phase of the properties of the medium. The hope
is that \Pb\ collisions at LHC energies (LHC energy is approximately
30 times higher than RHIC's maximum energy: LHC--energy/RHIC--energy $=$
5500/200 $=$ 27.5) will create a medium with an initial temperature
high enough so that the interaction weakens sufficiently to generate a 
a gaseous QGP.
\begin{acknowledgement}{\bf Acknowledgements}

  \noindent The author's research was supported by US Department of
  Energy, DE-AC02-98CH10886. The author would like to thank the
  organizers for an enjoyable and stimulating workshop. Special thanks
  to Mark Baker and Nu Xu for a careful reading of the manuscript and
  useful suggestions, and Ralf Averbeck for providing PHENIX data
  plotted on figure~\ref{fig7}.
\end{acknowledgement}

\begin{table}[ph]
  \caption{RHIC operating modes and total integrated luminosity delivered to all experiments for each run \cite{table}.}
\begin{center}
\begin{tabular}{|c|c|c|c|c|}\hline
\multirow{2}{*}{Run} &\multirow{2}{*}{Species}&\multirow{2}{*}{Total}& 
\multirow{2}{*}{Total}&\multirow{2}{*}{Average}\\
\multirow{2}{*}{} &\multirow{2}{*}{}&\multirow{2}{*}{particle}& 
\multirow{2}{*}{delivered}&\multirow{2}{*}{Store}\\
\multirow{2}{*}{} &\multirow{2}{*}{}&\multirow{2}{*}{energy: \sn}& \multirow{2}{*}{luminosity}&\multirow{2}{*}{polarization}\\
\multirow{2}{*}{} &\multirow{2}{*}{}&\multirow{2}{*}{[GeV/nucleon]}& \multirow{2}{*}{}&\multirow{2}{*}{}\\
\multirow{2}{*}{} &\multirow{2}{*}{}&\multirow{2}{*}{}& \multirow{2}{*}{}&\multirow{2}{*}{} \\ \hline
\multirow{2}{*}
{Run 1}          &Au+Au& 55.8  & $<$ 0.001 $\mu$b$^{-1}$  & - \\
                 &Au+Au& 130.4  & $<$ 20 $\mu$b$^{-1}$     & - \\
\hline
\multirow{3}{*}
                 &Au+Au    & 200.0  & $<$ 258 $\mu$b$^{-1}$  & - \\
{Run 2}          &Au+Au    & 19.6   & $<$ 0.4 $\mu$b$^{-1}$  & - \\
                 &pol. p+p & 200.0  & $<$ 1.4 pb$^{-1}$  & 15\% \\
\hline 
\multirow{2}{*}
{Run 3}          &d+Au     & 200.0  & 73 nb$^{-1}$  & - \\
                 &pol. p+p& 200.0  & $<$ 5.5 pb$^{-1}$     & 34\% \\
\hline
\multirow{3}{*}
              &Au+Au    & 200.0  & 3530 $\mu$b$^{-1}$  & - \\
{Run 4}       &Au+Au    & 62.4   & 67   $\mu$b$^{-1}$  & - \\
              &pol. p+p & 200.0  & 7.1 pb$^{-1}$  & 46\% \\
\hline 
\multirow{5}{*}
              &Cu+Cu   & 200.0  & 42.1 nb$^{-1}$  & - \\
              &Cu+Cu   & 62.4   & 1.5 nb$^{-1}$  & - \\
{Run 5}       &Cu+Cu   & 22.4   & 0.02 nb$^{-1}$  & - \\
              &pol. p+p& 200.0  & 29.5 pb$^{-1}$  & 46\% \\
              &pol. p+p& 409.8  & 0.1 pb$^{-1}$  & 30\% \\
\hline 
\multirow{2}{*}
{Run 6}       &pol. p+p& 200.0  & 88.6 pb$^{-1}$  & 58\% \\
              &pol. p+p& 62.4  & $<$ 1.05 pb$^{-1}$     & 50\% \\
\hline
\multirow{2}{*}
{Run 7}       &Au+Au   & 200.0  & 7250 $\mu$b$^{-1}$  & - \\
              &Au+Au   & 9.2  & test only & - \\
\hline
\multirow{3}{*}
              &d+Au    & 200.0  & 437 nb$^{-1}$  & - \\
{Run 8}       &pol. p+p& 200.0  & 38.4 pb$^{-1}$  & 45\% \\       
              &Au+ Au  & 9.2  &-  & - \\
\hline 
\end{tabular}
\end{center}
\label{tab1}
\end{table}

\begin{thebibliography}{99}
\bibitem{Shuryak80}S. Shuryak, Phys. Rep. \textbf{61}, (1980) 71;
L. McLerran, Rev. Mod. Phys. \textbf{58}, (1986) 1021.
\bibitem{Satz2003}H. Satz, Nucl. Phys. A \textbf{715} (2003) 3c.
\bibitem{Karsh2002}F. Karsch, Nucl. Phys. A \textbf{698} (2002) 199c.
\bibitem{Greiner1975}J. Hofman, {\it et al.}, in Bear Mountain Workshop, New York, December 19974;
H.G. Baumgardt, {\it et al.}, Z. Phys. A \textbf{273} (1975) 359.
\bibitem{table}http:$//$www.agsrhichome.bnl.gov$/$RHIC$/$Runs$/$
\bibitem{RHIC}I. Arsene \textit{et al.} Nucl. Phys. A \textbf{757} (2005) 1;
B. B. Back \textit{et al.} Nucl. Phys. A \textbf{757} (2005) 28;
J. Adams \textit{et al.} Nucl. Phys. A \textbf{757} (2005) 102;
K. Adcox \textit{et al.} Nucl. Phys. A \textbf{757} (2005) 184.
\bibitem{Rachid2003}R. Nouicer {\it et al.} , Eur. Phys. J. C \textbf{33} (2004) S606.
\bibitem{RachidMoriond2002}R. Nouicer {\it et al.}, The GIOI publishers : 2002 QCD and Hadronic Interactions, edited by Tran Thanh Van, (2002) 381.
\bibitem{AuAufrag}B.B. Back {\it et al.} Phys. Rev. Lett. \textbf{91}, (2003) 502303.
\bibitem{RachidPanic2006}R. Nouicer {\it et al.} AIP Conf. Proc. 842 (2006) 86; 
R. Nouicer e-Print: e-Print arXiv-nucl-ex/0601026;
\bibitem{CuCu2008}B. Alver \textit{et al.} e-Print: arXiv:0709.4008 (submitted to Phys. Rev. Lett. (2007)).
\bibitem{Dima}D. Kharzeev \textit{et al.} Nucl. Phys. A \textbf{747} (2005) 609; e-Prin arXiv:hep-ph/0408050.
\bibitem{RachidQM2004}R. Nouicer {\it et al.} J. Phys. G \textbf{30}, (2004) S113.
\bibitem{dAuPRL}B.B. Back {\it et al.} Phys. Rev. Lett. \textbf{93}, (2004) 082301.
\bibitem{dAuPRC}B.B. Back {\it et al.} Phys. Rev. C \textbf{72}, (2005) 031901(R). 
\bibitem{PHONIM}R. Nouicer {\it et al.} Nucl. Instrum. Meth. in Physics Research A \textbf{461}, (2001) 143; 
B.B. Back {\it et al.} Nucl. Instrum. Meth. in Physics Research A \textbf{499}, (2003) 603.
\bibitem{Thomas}T.S. Ullrich Eur. Phys. J. \textbf{A 19} (2004) s01. 
\bibitem{Hagedorn1965}R. Hagedorm Supl. A. Nuuvo Cimento Vol III \textbf{No.2} (1965) 150.
\bibitem{Munzinger}P. Braun-Munzinger {\it et al.}
  Phys. Lett. B \textbf{518} (2001) 41; e-Print arXiv-nucl-th/0304013
\bibitem{Brahms2003}I. G. Bearden {\it et al.} Phys. Rev. Lett. \textbf{90}, (2003) 102301.   
\bibitem{Star2008}J. Takahashi (for the STAR Collaboration), talk
  given at the International Conference on Strangeness in Quark Matter
  2008 (SQM2008), October 6, 2008 at Beijing, China. J. Takahashi 
  {\it et al.} e-Print arXiv[nucl-ex/0809.0823.
 
\bibitem{FlowSPS}C. Alt {\it et al.} Phys. Rev. C \textbf{68} (2003) 034903; and references therein.
\bibitem{hydro}T. Hirano , Acta Phys. Polon. B \textbf{36} (2005) 187; U.W. Heinz e-Print arXiv-nucl-th/0512051.  
\bibitem{NuXu}B.I. Abelev {\it et al.}  Phys. Rev. Lett. \textbf{99}
  (2007) 112301; e-Print arXiv-nucl-ex/0703033; N. Xu
  Braz. Jou. Phys. \textbf{37 2C} (2007) 773.
\bibitem{StarKprocess}J. Adams {\it et al.} Phys. Lett. B \textbf{612}, 181
 (2005) 181; S. Blyth  {\it et al.} proceedings
 of International Conference on Strangeness in Quark Matter,
 Los Angeles, California, 26 - 31 March 2006; e-Print arXiv-nucl-ex/060801
\bibitem{RHICStar}J. Adams \textit{et al.} Nucl. Phys. A \textbf{757} (2005) 102.
\bibitem{Huo}P. Huovinen {\it et al.} Phys. Lett. B \textbf{503} (2001) 58.   
\bibitem{Jacobs2005}P. Jacobs {\it et al.}, Prog. Nucl. Phys. 54, (2005) 443.
\bibitem{RachidQM2006}R. Nouicer {\it et al.} J. Phys. G \textbf{34} (2007) S887.
\bibitem{STARdi-jets}J. Adams {\it et al.} Phys. Rev. Lett. \textbf{91} (2003) 072304; 
J. Adams {\it et al.} Phys. Rev. Lett. \textbf{95} (2005) 152301; M.J. Tannenbaum PoS (CFRNC 2006) 001.  
\bibitem{BrahmsBaryon}I.G. Bearden {\it et al.}  Phys. Rev. Lett. \textbf{93} (2004) 102301;
\bibitem{Bjorken}J.D. Bjorken, Phys. Rev. D \textbf{27} (1983) 140;
  X.N. Wang {\it et al.} Phys. Rev. Lett. \textbf{68} (1992) 1480; and
  M. Gyulassy {\it et al.} Phys. Lett. B \textbf{243} (1990) 432.
\bibitem{Gaard}J.J. Gaardh$\phi$je {\it et al.} Nucl. Phys. A \textbf{734} (2004) 13.  
\bibitem{CompilationRaa}B.I. Abelev {\it et al.} Phys. Lett. B \textbf{655} (2007) 104; 
S.S. Adler {\it et al.} Phys. Rev. Lett. \textbf{98} (2007) 172302; 
S.S. Adler {\it et al.} Phys. Rev. Lett. \textbf{96} (2006) 202301; 
S.S. Adler {\it et al.} Phys. Rev. Lett. \textbf{91} (2003) 072303.
\bibitem{RAAdefinition}C. Albajar {\it et al.} Nucl. Phys. B \textbf{355} (1990)
  261; J. Adams {\it et al.} Phys. Rev. Lett. \textbf{91} (2003)
  172302.
\bibitem{Accardi}A. Accardi Contribution to the CERN Yellow report on
Hard Probes in Heavy Ion Collisions at the LHC (2002); e-Print:
arXiv:hep-ph/0212148.
\bibitem{Gyulassy2005}M. Gyulassy {\it et al.} Nucl. Phys. A \textbf{750} (2005) 30.    
\bibitem{Vitev1} I. Vitev {\it et al.} Nucl. Phys. A \textbf{715} (2003) 779;
  X.-N. Wang, Phys. Lett. B \textbf{595} (2004) 165; X.-N. Wang,
  Phys. Lett. B \textbf{579} (2004) 299; C.A. Salgado {\it et al.}
  Phys. Rev. D \textbf{68} (2003) 014008.
\bibitem{Vitev2} I. Vitev {\it et al.} Phys. Rev. Lett. \textbf{89} (2002) 252301.
\bibitem{Vitev3} I. Vitev, J. Phys. G \textbf{30} (2004) S791.
\bibitem{HeavyPHENIX}S.S. Alder {\it et al.}
  Phys. Rev. Lett. \textbf{98} (2007) 17230; S.S. Alder {\it et al.}
  Phys. Rev. Lett. \textbf{96} (2006) 032001; S.S. Alder {\it et al.}
  Phys. Rev. Lett.  \textbf{96} (2006) 032301;
\bibitem{HeavySTAR}J. Adams {\it et al.} Phys. Rev. Lett. \textbf{94} (2005) 062301; 
A.A.P. Suaide Braz. Jour. of Phys. \textbf{37} (2007) 2c. 
\bibitem{RachidNIM} M. Baker et al., Proposal for a Silicon Vertex
  Tracker (VTX) for the PHENIX Experiment, BNL- 72204-2004, Physics
  Dept. BNL (2004). R. Nouicer {\it et al.} Nuclear Instruments and
  Methods in Physics Research B \textbf{261} (2007) 1067; R. Nouicer
  {\it et al.}, POS VERTEX2007 (2007) 042.
\bibitem{STARNIM}C. Chasman {\it et al.} LBNL \textbf{5509} (2008).

%
%
\end{thebibliography}
\end{document}